\documentstyle[aps]{revtex}
\begin{document}
\title{Exact formulae for Higgs production through
       $\bbox{e\gamma\to e H}$ in the non-linear 
       $\bbox{R_{\xi}}$-gauge \\
          \vspace{-8ex}
           \hfill{\normalsize hep-ph/9704304 \\[1ex]}
           \hfill{\normalsize IFUAP-01-96\\}
          \vspace{2ex}}
\author{U. Cotti$^1$\protect\thanks{e-mail: ucotti@sirio.ifuap.buap.mx},
\,J.L. Diaz-Cruz$^1$\protect\thanks{e-mail: ldiaz@sirio.ifuap.buap.mx} 
and \, J.J. Toscano$^2$\protect\thanks{e-mail: jtoscano@fcfm.buap.mx} }
\address{$^1$Instituto de F\'{\i}sica, BUAP A.P. J-48, 72570 Puebla, 
Pue. M\'exico \\
$^2$Colegio de F\'{\i}sica, FCFM-BUAP, 72570 Puebla, Pue. M\'exico}

\date{April 11, 1997}
\maketitle

\begin{abstract}
We study the production of the SM Higgs boson ($H^0$) at future $e\gamma$ 
colliders, through the reaction $e\gamma\rightarrow eH^0$.
 The amplitude is evaluated using the non-linear $R_{\xi}$-gauge, which 
greatly simplifies the calculation. 
 Complete analytical expressions for the amplitudes are presented, which 
include the contributions from 1-loop triangles $\gamma\gamma^*H^0$ and
$\gamma Z^*H^0$ as well as the W- and Z-boxes with their related 
$eeH^0$ triangle graphs.
 The resulting cross section for this mechanism indicates that it 
could be used to detect the Higgs signal and to test its properties.
\end{abstract}
\pacs{13.85.Qk, 14.80.Er, 14.80.Gt}

\section{Introduction}
 The search for the standard Model (SM) Higgs boson at future colliders,
has become 
the focus of extensive studies, mainly because of its importance as a 
test of the mechanism of electroweak symmetry breaking~\cite{kane:90}. 
 At the {\sl next} $e^+e^-$ {\sl linear collider} (NLC), it will be 
possible to study some features of the SM Higgs boson ($H^0$),
through the 
production reactions $e^+e^- \rightarrow H Z$, 
and $e^+e^- \rightarrow H \gamma$~\cite{npb267:509,prd49:1265,%
prd53:3616}.
 However, given the prospect for studying $e\gamma$ collisions at the 
NLC through its operation in the photon mode~\cite{rnc16:1}, 
it is convenient to have a production mechanism that can be used to 
study the detection of the Higgs boson in this mode. 
 Besides the interest to detect a Higgs boson in any machine, 
it is important to study carefully the possibility
to test its couplings to all the SM particles, since this 
will help to determine whether it corresponds to the SM or to 
some of its extensions.

 The production of $H^0$ in $e \gamma$ has already been evaluated using 
the Williams-Weizsacker approximation, namely through the 
$\gamma\gamma \rightarrow H$ mechanism with 
on-shell photons~\cite{npb421:65,prd48:1430,prd49:91,prd50:3546}, 
with the conclusion that this reaction 
will dominate over the mode $e^+ e^- \rightarrow H \gamma$, 
and it can be usefull to study the vertex $H \gamma \gamma$.
 However, in order to improve the precision of the calculation,
and the confidence on the result, it is necessary
to consider the full two-body reaction 
$e \gamma \rightarrow e H^0$, which includes the contributions
from 1-loop vertices $\gamma \gamma^* H^0$, $\gamma Z^* H^0$,
and the W- and Z-boxes. 
 The reaction $e \gamma \rightarrow e H^0$
can be used also to study the vertex 
$H \gamma Z$, provided that it could be possible to separate its 
effects from the other contributions. 
 
 In this paper, we calculate the cross-section for the reaction 
$e \gamma \rightarrow H e$, using a nonlinear 
$R_{\xi}$-gauge~\cite{prd6:2923,ap94:349,npb193:257,npb213:390}.
 We present complete analytical expressions for the amplitudes, and 
show the power of the non-linear gauge in verifying the gauge 
invariance of the result.
In this gauge some 3-point vertices $WVG$, 
as well as some 4-point vertices $WVGH$, are absent;
$V$ represents the neutral gauge bosons
$Z, A$ and $G^\pm$ denotes the charged Goldstone boson.
Thus, the number of diagrams is reduced considerably.

 In order to derive the Feynman rules arising from the nonlinear 
$R_{\xi}$ gauge, one needs to specify the gauge fixing term, which 
has the form \cite{prd49:1265}
\begin{equation}
 {\cal L}_{GF} = -\frac{1}{2\xi_Y}
 \left(
  f^Y
 \right)^2 - \frac{1}{2\xi_i}
 \left(
  f^i
 \right)^2,
\end{equation}
where the $\rm SU(2)_L$ and $\rm U(1)_Y$ $f$-functions are given by:
\begin{equation}
 f^i = 
 \left[ 
  \delta^{ij} \partial_\mu - g^\prime B_\mu \epsilon^{ij3}
 \right] 
 W^{j \mu} + ig\xi_i 
 \left[
  \phi'^\dagger \frac{\tau^i}{2} \langle \phi \rangle_0 -
  \langle \phi \rangle^\dagger_0 \frac{\tau^i}{2} \phi' +
  i \epsilon^{ij3} \phi'^\dagger\frac{\tau^j}{2} \phi'
 \right],
\end{equation}
\begin{equation}
 f^Y = \partial_\mu B^\mu + ig' \xi_Y 
 \left[ 
  \phi'^\dagger \frac{1}{2} \langle \phi \rangle_0 -
  \langle \phi \rangle^\dagger_0 \frac{1}{2} \phi'
 \right],
\end{equation}
respectively, with 
$\langle \phi \rangle^\dagger_0 = \frac{1}{\sqrt{2}}(0,v)$ 
denoting the v.e.v. of the neutral component of the Higgs 
doublet and $\phi'= \phi- \langle \phi \rangle_0$.
 Using now the definitions: $ f^\pm = \frac{f^1 \mp if^2}{\sqrt{2}}$,
$f^Z = {\rm c}_W f^3 - {\rm s}_W f^Y$, 
$f^A = {\rm c}_W f^3 + {\rm s}_W f^Y$, 
${\rm c}_W = \cos \theta_W$, ${\rm s}_W = \sin \theta_W$,
one obtains:
\begin{mathletters}
\begin{eqnarray}
 f^+ &=& 
  \left( 
   \partial_\mu - i g' B_\mu 
  \right) W^{+\mu }+ ig 
  \left( 
   \xi_1 - \xi_2
  \right)
  G^-_W \frac{(v+H^0+iG_Z)}{4} - ig   
  \left(
   \xi_1 + \xi_2
  \right) G^+_W \frac{(v+H^0-iG_Z)}{4}, \\
 f^Z &=& \partial_\mu Z^\mu - 
 \left( 
  {\rm c}^2_W \xi_3 + {\rm s}^2_W \xi_Y
 \right) M_Z G_Z , \\
 f^A &=& \partial_\mu A^\mu - \frac{{\rm s}_{2W}}{2}
 \left(
  \xi_3- \xi_Y
 \right) M_Z G_Z .
\end{eqnarray}
\end{mathletters}

 The gauge parameters $\xi_Y, \xi_i$ are all chosen to be unity,
which defines the non-linear 't~Hooft-Feynman gauge. 
 After substituting these expressions in the gauge-fixing lagrangian, 
and including them in the full gauge and Higgs lagrangian, one derives 
the non-linear vertices for the scalar and 
gauge-sector, which are summarized in ref.~\cite{prd49:1265}.

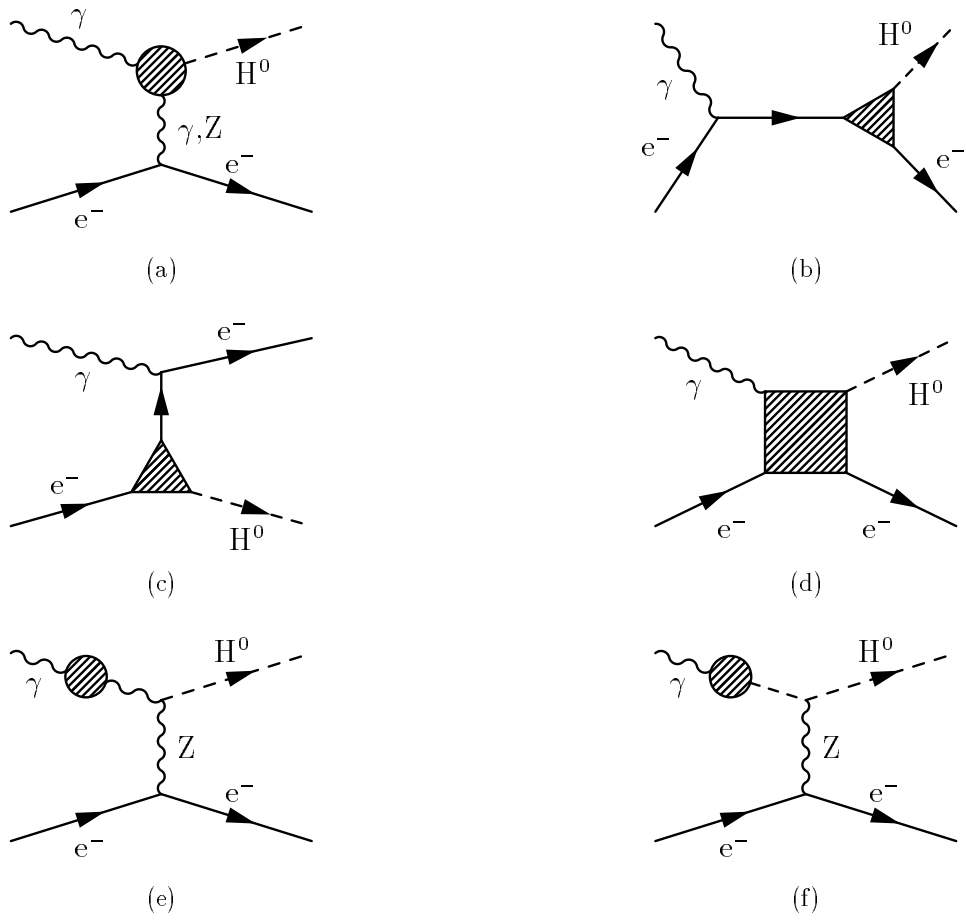
\begin{figure}
 \caption{Classification of graphs that contributes to the 
          reaction $e \gamma \rightarrow e H^0$.%
           \label{diagr}
          }
\end{figure}

 The diagrams encountered in the calculation of the 
$e^- \gamma \rightarrow e^- H^0 $ include: i) graphs with the triangle 
$AA^*H$ and $AZ^*H$ loop (Fig.~\ref{diagr}a), ii) graphs with 
$W-$ and $Z-$mediated box diagrams and the related triangles with 
external fermion legs (Figs.~\ref{diagr}b, \ref{diagr}c, \ref{diagr}d),
and ii) those with $Z$-$A$ and $Z$-$H$ self energies  
(Figs.~\ref{diagr}e, \ref{diagr}f). It results that these sets are 
{\em separately} gauge invariant. 

 The diagrams of Figs.~\ref{diagr}a involves a virtual gauge
bosons ($Z^0$ or $\gamma$) in the t-channel.
 The relevant contributions to these triangle graphs include the 
heaviest fermions (the top and bottom quarks), $W^\pm$-$G^\pm$ bosons
and ghosts, however the amplitudes for each subset are also separately gauge 
invariant, thus the total amplitude is also gauge invariant.
The remaining diagrams (Figs.~\ref{diagr}b, \ref{diagr}c, \ref{diagr}d)
have no gauge boson poles. 
 They consist of box diagrams, with the Higgs boson emerging from one 
of the box vertices, together with associated triangle diagrams $eeH^0$. 
There are two such combinations of boxes and triangles: one with 
$Z$'s in the loops and one with $W$'s in the loops.
The amplitude for the graphs of Fig. 1-e, which consist of 
tadpole and bubble diagrams, with fermions, $W^\pm$-$G^\pm$ 
bosons and ghosts in the loops, combine to give vanishing results,
which is a consequence of using the non-linear gauge.
On the other hand, the amplitude for the graphs of
Fig. 1-f vanish in the approximation $m_e=0$.

 We have evaluated the amplitudes using dimensional
regularization, with the help of the programs 
Feyncalc~\cite{cpc64:345} and the numerical package 
FF~\cite{zpc46:425,cpc66:1}.
Our result for the total amplitude is written as follows:
\begin{equation}
 {\cal M} = {\cal M}_{\gamma} + {\cal M}_Z + {\cal M}^{\rm box}_{Z} 
            + {\cal M}^{\rm box}_W .
\end{equation}

 The contribution to the matrix element 
from the $A-$ and $Z-$triangles is given by:
\begin{equation}
 {\cal M}_{\gamma, Z} = 
  \frac{i \alpha^2 M_W}{4 {\rm s}^3_W{\rm c}^4_W}
  \overline{u}(p_2) (a_{\gamma,Z}-b_{\gamma,Z} \gamma_5) \gamma^\nu u(p_1)
  \epsilon^\mu (k_1, \lambda_1)
  F_{\gamma,Z} 
  \left( 
   k_1 \cdot k_2 g_{\mu \nu} - k_{1\nu} k_{2\mu}
  \right) ,
\end{equation}
where
\begin{eqnarray}
 F_\gamma  &=& 
 \frac{4 {\rm s}^2_W {\rm c}^4_W}{M_W^2 t}
 \left\{
  2 N_C Q_f^2 H_f + \frac{\lambda_W}{\lambda_W - \tau_W}
  \left\{
    - \frac{1}{2} \frac{\tau_W}{\tau_W - \lambda_W}
   (2 + 7\tau_W)
   \left[
    B_0
    \left(
     M_H^2, M_W^2, M_W^2
    \right)
    - B_0
    \left(
     t, M_W^2, M_W^2
    \right)    
   \right]
    \right. \right. \nonumber \\[2ex]
   && \left. \left.
   \hspace{12ex} +2 + 3 \tau_W +
   \left(
    2 + 3 \tau_W + 8 \frac{\tau_W - \lambda_W}{\lambda_W}
   \right)
   2 M_W^2 C_0
   \left(
    t, M_H^2, M_f^2 
   \right)   
  \right\}
 \right\},
\end{eqnarray}
\begin{eqnarray}
 F_Z  &=& 
 \frac{{\rm c}^4_W}{ M_W^2(t-M^2_Z)}
 \left\{
 - \frac{C_V^f N_C Q_f H_f}{{\rm c}^2_W} 
  + \; 4 (3 - {\rm t}^2_W) M_W^2 C_0
  \left(
   t, M_H^2, M_W^2 
  \right)
  \right. \nonumber \\[2ex]
  && \left.
  + \; \frac{\tau_W \lambda_W}{2(\tau_W - \lambda_W)}
  \left[
   5 + \frac{2}{\tau_W} - 
   \left(
    1 + \frac{2}{\tau_W}
   \right)
   {\rm t}^2_W
  \right]
  \left\{
   1 - \frac{\tau_W}{2}
   \left[
    B_0
    \left(
     M_H^2, M_W^2, M_W^2
    \right)
    - B_0
    \left(
     t, M_W^2, M_W^2
    \right)    
   \right]
    \right. \right. \nonumber \\[2ex]
   && \left. \left.
   \hspace{12ex} 
   + \; 2 M_W^2 C_0
   \left(
    t, M_H^2, M_W^2 
   \right)   
  \right\}
 \right\}.
\end{eqnarray}
The function $H_f$ represents the fermionic contribution to 
the loops and is given by:
\begin{eqnarray}
 H_f&=&
   \frac{\tau_f \lambda_f}{\tau_f - \lambda_f} 
   - \frac{\lambda_f \tau_f^2}{(\tau_f - \lambda_f)^2}
   \left[
    B_0
    \left(
     M_H^2, M_f^2, M_f^2
    \right)
    - B_0
    \left(
     t, M_f^2, M_f^2
    \right)
   \right] 
   + 2
   \left(
    1 + \frac{\tau_f \lambda_f}{\tau_f - \lambda_f}
   \right)
   M_f^2 C_0
   \left(
    t, M_H^2, M_f^2 
   \right),
\end{eqnarray}
with 
\begin{eqnarray}
 & a_\gamma = 1,       \hspace{2ex} 
   b_\gamma = 0,       \hspace{2ex} 
   a_Z = 1 - 4 {\rm s}^2_W, \hspace{2ex} 
   b_Z = 1,            \hspace{2ex}
 & \nonumber \\
 & \tau_x = \frac{4 M_x^2}{M_H^2}, \hspace{3ex}
  \lambda_x = \frac{4 M_x^2}{t},   \hspace{3ex}  
  {\rm t}_x = \frac{{\rm s}_x}{{\rm c}_x}.
 &
\end{eqnarray}

 Although one can write analytical expressions for the previous
$B_0$ and $C_0$ scalar functions, they are also evaluated with the  
help of the FF package. The full dependence of the above $C_0$'s is
the following:
\begin{mathletters}
\begin{eqnarray}
 C_{0}(t,M^2_H,M^2_t)&=& C_0(0,t,M^2_H,M^2_t,M^2_t,M^2_t), \\
 C_{0}(t,M^2_H,M^2_W)&=& C_0(0,t,M^2_H,M^2_W,M^2_W,M^2_W).	
\end{eqnarray}
\end{mathletters}

 The result for the contribution of the Z-mediated box diagram
to the amplitude, including the related $Hee$ triangle graph,
is the following:

\begin{eqnarray}
 {\cal M}^{\rm box}_Z &=& 
  \frac{i \alpha^2 M_Z}{4 {\rm s}^3_W {\rm c}^3_W}
  \overline{u}(p_2) \gamma^\nu (a_Z - \gamma_5)^2 u(p_1)
  \epsilon^\mu (k_1, \lambda_1)
  \left[
   - A (t, s, u) 
   \left(
    k_1 \cdot p_1 g_{\mu \nu} - p_{1\mu} k_{1\nu} 
   \right)
   \right. \nonumber \\[2ex]
   & & \left.
   \hspace{42ex} 
   + \; A (t, u, s) 
   \left(
    k_1 \cdot p_2 g_{\mu \nu} - p_{2\mu} k_{1\nu}
   \right)
  \right],
\end{eqnarray}
with $s=(p_1+k_1)^2, \, t=(k_2-k_1)^2, \, u=(k_1-p_2)^2$, and
\begin{eqnarray} 
 A(t, s, u) &=&
 \frac{1}{2st}
 \left[
  \left(
   \frac{s - M_Z^2}{s}
  \right)
  \left(
   M_Z^2 (s + u) - su 
  \right)    
  D_{0Z}(1, 2, 3, 4) 
  \right. \nonumber \\[2ex]
  && \left.
  \hspace{5ex}
  + \;
  \left(
   s - M_Z^2
  \right)
   \left[
    C_{0Z}(1, 2, 4) + \frac{u}{s} C_{0Z}(1, 2, 3) 
    - \frac{t+u}{s} C_{0Z}(2, 3, 4)
    \right. \right. \nonumber \\[2ex]
    && \left. \left.
    \hspace{4ex}
    + \; \frac{1}{s}  
    \left(
     t -s -2M_Z^2 \frac{st}{(s +t)(s -M_Z^2)}
    \right)
   C_{0Z} (1, 3, 4)
  \right]
  + \;  \frac{2t}{s + t}
  \left(
   B_{0Z}(3, 4) - B_{0Z}(1, 3)
  \right)  
 \right],
\end{eqnarray}
and the arguments of the scalar functions are
\begin{mathletters}
\begin{eqnarray}
 B_{0Z}(1,3)     &=& B_0 (u,   , m_e^2, M_Z^2), \\[2ex]
 B_{0Z}(3,4)     &=& B_0 (M^2_H, M_Z^2, M_Z^2), \\[2ex]
 C_{0Z}(1,2,3)   &=& C_0 (0,     0,     u, m_e^2, m_e^2, M_Z^2), \\[2ex]
 C_{0Z}(1,2,4)   &=& C_0 (0,     s,     0, m_e^2, m_e^2, M_Z^2), \\[2ex]
 C_{0Z}(1,3,4)   &=& C_0 (0, M^2_H,     u, m_e^2, M_Z^2, M_Z^2), \\[2ex]  
 C_{0Z}(2,3,4)   &=& C_0 (s,     0, M^2_H, M_Z^2, m_e^2, M_Z^2), \\[2ex]
 D_{0Z}(1,2,3,4) &=& D_0 (0,s,M^2_H,u,0,0,m^2_e,m^2_e,M^2_Z,M^2_Z).
\end{eqnarray}
\end{mathletters}
 The labels 1, 2, 3, 4 in $D_0$ are associated to the internal loop-masses,
namely to the last four entries of $D_0$. Then, 
the $B_0 (C_0) $ functions are obtained by
suppressing two (one) of the propagators that appear in the
$D_0$ function.

 The expression for the amplitude coming from the
W-box and its related $eeH^0$ triangle graph
is given by:
\begin{eqnarray}
 {\cal M}^{\rm box}_W &=& 
  \frac{i \alpha^2 M_W}{2 {\rm s}^3_W}
  \overline{u}(p_2) \gamma^\nu (1 - \gamma_5)^2 u(p_1)
  \epsilon^\mu (k_1, \lambda_1) 
  \left[
   \left(
    A_1 (t, s, u) 
    + \; A_2 (t, u, s)
   \right)
   \left(
    k_1 \cdot p_1 g_{\mu \nu} - p_{1\mu} k_{1\nu} 
   \right)
   \right. \nonumber \\[2ex]
   & & \left.
   \hspace{42ex} 
   - \;
   \left(
     A_2 (t, s, u) +  A_1 (t, u, s)
   \right)
   \left(
    k_1 \cdot p_2 g_{\mu \nu} - p_{2\mu} k_{1\nu}
   \right)
  \right],
\end{eqnarray}
where
\begin{eqnarray} 
 A_1 (t, s, u) &=&
 \frac{1}{2st}
 \left[
  \left(
   \frac{s - M_W^2}{s}
  \right)
  \left(
   M_W^2 (s + u) + st 
  \right)  
  D_{0W}(1, 2, 3, 4) 
  \right. \nonumber \\[2ex]
  && \left.
  \hspace{5ex}
  + \;
  \left(
   s - M_W^2
  \right)
   \left[
    C_{0W}(2, 3, 4) - \frac{t}{s} C_{0W}(1, 3, 4) 
    + \frac{u + t}{s} C_{0W}(1, 2, 4)
   - \frac{s + u}{s} C_{0W} (1, 2, 3)
  \right] \right. \nonumber \\[2ex]
  && \left.
  \hspace{5ex}  
  + \frac{2t}{s + u}
  \left[
   B_{0W}(1, 2) - B_{0W}(1, 3)
  \right]
 \right],
\end{eqnarray}
\begin{eqnarray} 
 A_2 (t, s, u) &=&
 \frac{1}{2tu}
 \left[
  \left(
   \left(
    \frac{t + u - M_W^2}{u}
   \right)
   \left(
    M_W^2 (s + u) - st 
   \right)
   - 2 M_W^2 t
  \right)  
  D_{0W}(1, 2, 3, 4) 
  \right. \nonumber \\[2ex]
  && \left.
  \hspace{5ex}
  + \;
  \left(
   t + u - M_W^2
  \right)
   \left[
    \frac{s}{u} C_{0W}(2, 3, 4) + \frac{t}{u} C_{0W}(1, 3, 4) 
    - \frac{s + u}{u} C_{0W}(1, 2, 3)
   + \frac{u^2 - 2tu - t^2}{u (t + u)} C_{0W}(1, 2, 4)
  \right] \right. \nonumber \\[2ex]
  && \left.
  \hspace{5ex}  
   + \;\frac{2t}{t + u}
  \left[
   B_{0W}(2, 4) - B_{0W}(1, 2)
  \right]
  + \frac{2t}{s + u}
  \left[
   B_{0W}(1, 3) - B_{0W}(1, 2)
  \right]
 \right].
\end{eqnarray}

 The scalar functions that appear before have the following
arguments:
\begin{mathletters}
\begin{eqnarray}
 B_{0W}(1,2)     &=& B_0 (M^2_H, M^2_W, M^2_W), \\[2ex]
 B_{0W}(1,3)     &=& B_0 (t,     M^2_H, M^2_H), \\[2ex]
 B_{0W}(2,4)     &=& B_0 (s,         0, M^2_W), \\[2ex]
 C_{0W}(1,2,3)   &=& C_0 (t, 0, M^2_H, M^2_W, M^2_W, M^2_W), \\[2ex]
 C_{0W}(1,2,4)   &=& C_0 (0, s, M^2_H, M^2_W,     0, M^2_W), \\[2ex]
 C_{0W}(1,3,4)   &=& C_0 (0, 0,     t, M^2_W,     0, M^2_W), \\[2ex]  
 C_{0W}(2,3,4)   &=& C_0 (0, 0,     s,     0, M^2_W, M^2_W), \\[2ex]
 D_{0W}(1,2,3,4) &=& D_0 (0, 0, 0, M^2_H, t, s, M^2_W, 0, M^2_W, M^2_W). 
\end{eqnarray}
\end{mathletters}

 In order to obtain the cross-section, one has to square the 
total amplitude, sum and average over initial and final 
polarizations, then the correspondent differential cross section for the
reaction $\gamma + e \rightarrow H + e$ is expressed as follows:

\begin{equation}
 \frac{ d \widehat{\sigma}}{dt} = 
 \frac{1}{16 \pi s^2} |\overline{\cal M}|^2
\end{equation}
where:
\begin{eqnarray}
 | \overline{\cal M}|^2 &=& 
 \frac{\alpha^4 M^2_W (-t)}{64{\rm s}^6_W{\rm c}^8_W}
 \left\{
  \left(
   s^2+u^2
  \right)
  \left[
   \left|
    F_{\gamma}
   \right|^2 
   + 2a_Z Re(F^*_\gamma F_Z) + (1 + a_Z^2)
   \left|
    F_Z
   \right|^2 
  \right]
   + s^2 f_s + u^2f_u
 \right\},
\end{eqnarray}
with 
\begin{eqnarray} 
 f_s &=& 
 |A_{12}|^2 + (1 + 6 a_Z^2 + a_Z^4) |A_s|^2 
 + 2 Re(A_{12} F^*_\gamma)
 -2 (1 + a_Z^2) Re(A_s F^*_\gamma) \nonumber \\
 &&  
 -2(1 + a_Z)^2 Re(A_{12} A^*_s) + 4(1 + a_Z) Re(A_{12} F^*_Z)
 -2(a_Z^3 + 3a_Z) Re(A_s F^*_Z),
\end{eqnarray}
with $f_u = f_s( s \to u; A_{12} \to -A_{21}; A_s \to -A_u)$,
$A_s = A(t,s,u), \, A_u = A(t,u,s)$, and also:
\begin{eqnarray} 
 A_{12}&=& 2{\rm c}^4_W [ A_1(t,s,u)+A_2(t,u,s)]   \nonumber \\
 A_{21}&=& 2{\rm c}^4_W [ A_2(t,s,u)+A_1(t,u,s)]. 
\end{eqnarray}

 Finally, in order to obtain the total
cross-section ($\sigma_T$), one needs to convolute
$\widehat{\sigma}$ with the photon distribution 
\cite{prd49:91}, namely

\begin{equation} 
 \sigma_T = \frac{1}{S} \int^{0.83S}_{M^2_H} F_{\gamma}( \frac{s}{S})
             \hat{\sigma}(s),
\end{equation}
where $S$ denotes the squared c.m. energy of the $e^+e^-$-system, and the
photon distribution is given by

\begin{equation} 
 F_\gamma(x)=\frac{1}{D(\xi)} 
 \left[ 
  1-x+\frac{1}{1-x}\frac{4x}{\xi(1-x)}-\frac{4x^2}{\xi^2(1-x)^2}
 \right],
\end{equation}
where:
\begin{equation} 
 D(\xi)=
 \left[ 
  (1-\frac{4}{\xi}-\frac{8}{\xi^2}) \log(1+\xi)
  +\frac{1}{2}+\frac{8}{\xi}-\frac{1}{2(1+\xi)^2}
 \right].
\end{equation}

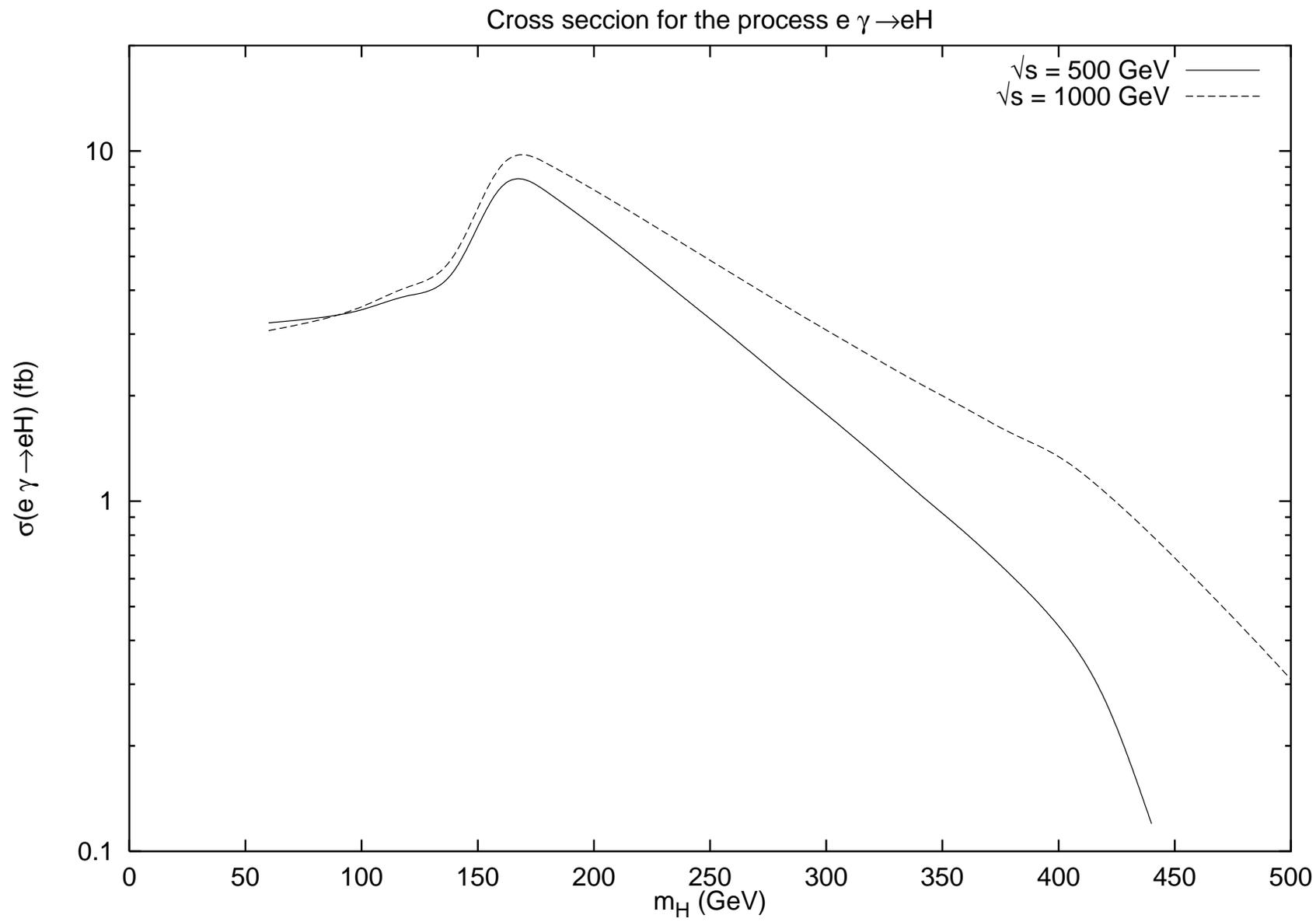
\begin{figure}
 \caption{The cross section for the reaction 
          $e \gamma \rightarrow e H^0$ at 
          $\protect\sqrt{s}$ = 500 GeV and $\protect\sqrt{s}$ = 1 TeV.}
 \label{plot}
\end{figure}

 We present in Fig.~\ref{plot} the resulting cross-section for
our process.
We have taken $m_t\sim 175\;$ GeV~\cite{prl73:225}, 
and the values for the electroweak parameters given in the
table of particle properties \cite{prd54:1}.
In fact, it happens that the contributions of the
boxes can be neglected.
Although the cross section is small, about 6 fb for $m_H=200$
GeV and $E_{c.m.}=500$ GeV,
it is larger than the one for the reaction 
$e^+e^- \to H+\gamma$ \cite{prd53:3616}. 
With the expected NLC luminosities (100 fb$^{-1}$/yr)
it will be possible to observe up to
about 600 $H + e$ events, which should allow to
study the properties of the Higgs boson. 
Our results reproduce correctly
the value reported in Ref. \cite{prd53:3616}
which uses only the photon-pole contribution, 
for the total cross-section;
we find $\sigma=6.1$ fb, whereas
they find $\sigma=5.9$ fb, for $E_{c.m.}=500$ GeV and
$M_H=200$ GeV.

 The final signature depends on the Higgs boson mass.
For instance, if we focus on the intermediate-mass region
($M_W < M_H < 2M_W$) the dominant Higgs decay is into
$b\bar b$ pairs, and in order to estimate the backgrounds
that need to be considered,
one could rely on the results of Ref.~\cite{prd53:3616},
which considers the production of $e\gamma\to e+b\bar{b}$
in the context of their study of the production
of the pseudoscalar $A^0$ of the MSSM, and they conclude that
this background can be handled, and detection of the signal is
possible. In our case, since the event rates are of the same order,
we believe that the signal from $H^0$ can be detected too
\footnote{ After completion of this work, we became aware that
the same calculation has been performed by E. Gabrielli et al.
\cite{hep-ph:9702414}, but in the linear gauge. Comparing
the number of graphs arising in each method illustrates
the power of the non-liner gauge, 
since the number of graphs that
we have evaluated is reduced considerably. 
Some differences among our works can be noticed,
for instance in our paper the cross-section includes
folding with the photon distributions, which  
is not done in their paper. 
We also agree with Ref.\cite{prd53:3616}, 
that the Williams-Weizsacker approximation
overestimates the exact result, whereas in ref. \cite{hep-ph:9702414} 
it is found the opposite.}.

In conclusion, we have studied the production of the SM Higgs boson 
($H^0$) at future $e\gamma$ 
colliders, through the reaction $e\gamma\rightarrow eH^0$.
 The amplitude is evaluated using a non-linear $R_{\xi}$-gauge, which 
greatly simplifies the calculation, and allows to present 
complete analytical expressions for the amplitudes.
 The resulting cross section for this mechanism indicates that it could 
 allow the detection of the Higgs boson, and can also be 
used to test the Higgs boson properties.

\acknowledgments We acknowledge financial support from
CONACYT and SNI (Mexico).

\def\href#1#2{#2} 
\begingroup\raggedright\endgroup

\end{document}